\documentclass[aps,prl,twocolumn,amsmath,amssymb,superscriptaddress,floatfix]{revtex4-2}

\usepackage{graphicx}
\usepackage{bm}
\usepackage{color}
\usepackage[normalem]{ulem}
\usepackage{txfonts}
\usepackage[colorlinks=true, urlcolor=blue, linkcolor=blue, citecolor=blue]{hyperref}
\usepackage{physics}

\begin{document}

\title{Extracting Universal Entanglement Scaling from Mixed Fermionic  Gaussian States via Entanglement Projected Entropy}

\author{Jia-Wen Tao}
\affiliation{School of Physical Science and Technology, ShanghaiTech University, Shanghai 201210, China}

\author{Hui-Ke Jin}
\email{jinhk@shanghaitech.edu.cn}
\affiliation{School of Physical Science and Technology, ShanghaiTech University, Shanghai 201210, China}
\date{\today}

\begin{abstract}
Identifying spatial quantum correlations in mixed states is challenging because thermal mixed-state contributions obscure the entanglement encoded in subsystem entropy.
Here, we introduce the entanglement projected entropy (EPE), a purification-independent Gaussian spatial filter for mixed fermionic states.
By resolving subsystem entropy into Gaussian entropy  channels and projecting their purification partners onto the physical complement, we obtain a closed-form expression in terms of the physical covariance matrix.
In a one-dimensional free-fermion chain, it removes the volume-law mixed-state background and recovers the zero-temperature conformal scaling with the $c/3$ coefficient. In a two-dimensional half-filled $\pi$-flux model, it reveals a universal finite-temperature scaling collapse governed by a Dirac infrared length fixed by the low-energy velocity. 
These results establish EPE as an entropy-channel filter that exposes boundary-sensitive universal scaling hidden beneath mixed-state entropy.
\end{abstract}
\maketitle

{\em Introduction.---}
Quantum entanglement provides a central organizing principle in modern many-body physics~\cite{Amico2008, Calabrese2004,EisertCramerPlenio2010}, with applications ranging from quantum criticality~\cite{Vidal2003, Calabrese2004, Calabrese2009} and topological phases~\cite{Kitaev2006, Levin2006,Chen2010} to holography~\cite{Ryu2006} and quantum information~\cite{NielsenChuang}. For a pure state $|\Psi\rangle$, the bipartite von Neumann entropy (vNE)
\begin{equation}
    \mathcal S(A)=-\Tr_A(\rho_A\ln \rho_A),
    \qquad
    \rho_A=\Tr_B |\Psi\rangle\langle\Psi|,
\end{equation}
provides a canonical measure of spatial entanglement across a bipartition $S=A\oplus B$, a fundamental concept tracing back to the earliest formulation of quantum subsystems. Its scaling structure has become a standard tool for diagnosing universal boundary-law signatures~\cite{Bombelli1986, Srednicki1993, EisertCramerPlenio2010,Wolf2006, GioevKlich2006, Swingle2010}, including the logarithmic scaling of conformal field theories (CFT)~\cite{DiFrancesco1997, Holzhey1994, Calabrese2004} and the topological entanglement entropy in gapped phases~\cite{Kitaev2006, Levin2006}.

However, in mixed states the vNE $\mathcal S(A)$ is no longer an unambiguous measure of spatial entanglement, since it also contains thermal and classical contributions~\cite{Calabrese2004, Wolf2008,PeschelEisler2009,HertzbergWilczek2011}. In such settings, $\mathcal S(A)$ contains a thermodynamic contribution typically scaling as $\propto|A|$, which can overwhelm the boundary-law or other subextensive terms associated with spatial entanglement.

This ambiguity has motivated the development of a variety of mixed-state diagnostics~\cite{Horodecki2009}.
Prominent examples include mutual information for total correlations~\cite{Wolf2008,Henderson2001,Iaconis2013}, logarithmic negativity and partial-transpose constructions~\cite{Vidal2002,Plenio2005,Calabrese2012,Calabrese2015,Eisler2015,Coser2015,Ruggiero2016,Shapourian2017,Shapourian2019}, purification-based quantities such as the entanglement of purification and reflected entropy~\cite{Terhal2002,Nguyen2018,Takayanagi2018,Bueno2020,Dutta2019}, and squashed entanglement~\cite{Tucci2002,Christandl2004}. 
This variety reflects that no single diagnostic is optimal for all mixed-state problems. In practice, the useful diagnostic depends on the correlations of interest and on the structure of the state.

In this Letter, we introduce the \emph{entanglement projected entropy} (EPE), a quantity designed to isolate boundary-sensitive quantum correlations across a spatial bipartition from extensive mixed-state backgrounds. 
The key idea is to resolve $\mathcal S(A)$ into Gaussian entropy channels and weight each channel by the support of its purification partner in the physical complement $B$. This yields a purification-independent expression entirely in terms of the physical covariance matrix.
Within this framework, we demonstrate two universal scaling results. In a one-dimensional (1D) critical free-fermion system at finite temperature, EPE removes the extensive mixed-state contribution and analytically recovers the zero-temperature logarithmic scaling governed by CFT. In a two-dimensional (2D) half-filled $\pi$-flux model~\cite{Affleck1988}, it reveals a finite-temperature scaling collapse controlled by an effective Dirac infrared length.

{\em Gaussian formulation of EPE.---} 
We begin with a $U(1)$-symmetric mixed fermionic Gaussian state $\rho_S$ on a physical system $S$~\cite{Bravyi2004,Peschel2003, PeschelEisler2009}. As a free-fermion state, $\rho_S$ is fully characterized by its centered covariance matrix
\begin{equation}
    C_{jl}=\Tr\!\left(\rho_S\,[a_l^\dagger,a_j]\right)
    =2\langle a_l^\dagger a_j\rangle-\delta_{jl},\label{eq:C}
\end{equation}
where $a^\dagger_j$ denotes the fermion creation operator at site $j$. Note that for a pure Gaussian state, one has $C^2=I$, whereas for a mixed state the spectrum of $C$ lies in the interval $[-1,1]$.

Consider a bipartition $S=A\oplus B$. The reduced state on $A$ is also Gaussian and fully determined by its restricted covariance matrix $C_A=P_A C P_A,$ where $P_A$ is the projector onto subsystem $A$. The vNE of subsystem $A$ then is~\cite{Peschel2003,PeschelEisler2009} 
\begin{equation}
    \mathcal{S}(A)=\sum_{q}s(\lambda_q) = \Tr_A[s(C_A)], \label{eq:vNE_gaussian}
\end{equation}
where $s(\lambda)=-\frac{1+\lambda}{2}\ln\frac{1+\lambda}{2}-\frac{1-\lambda}{2}\ln\frac{1-\lambda}{2},$ and $\lambda_q$ are the eigenvalues of $C_A$.
When $\rho_S$ is pure, $\mathcal{S}(A)$ faithfully captures the boundary-law entanglement structure~\cite{Peschel2003,PeschelEisler2009,Wolf2006,GioevKlich2006,Swingle2010}. For mixed states, however, it sums all contributions to the local entropy, including those not associated with spatial entanglement across the $A$-$B$ cut~\cite{PeschelEisler2009,HertzbergWilczek2011}.

\begin{figure}[!t]
    \centering
    \includegraphics[width=1.\linewidth]{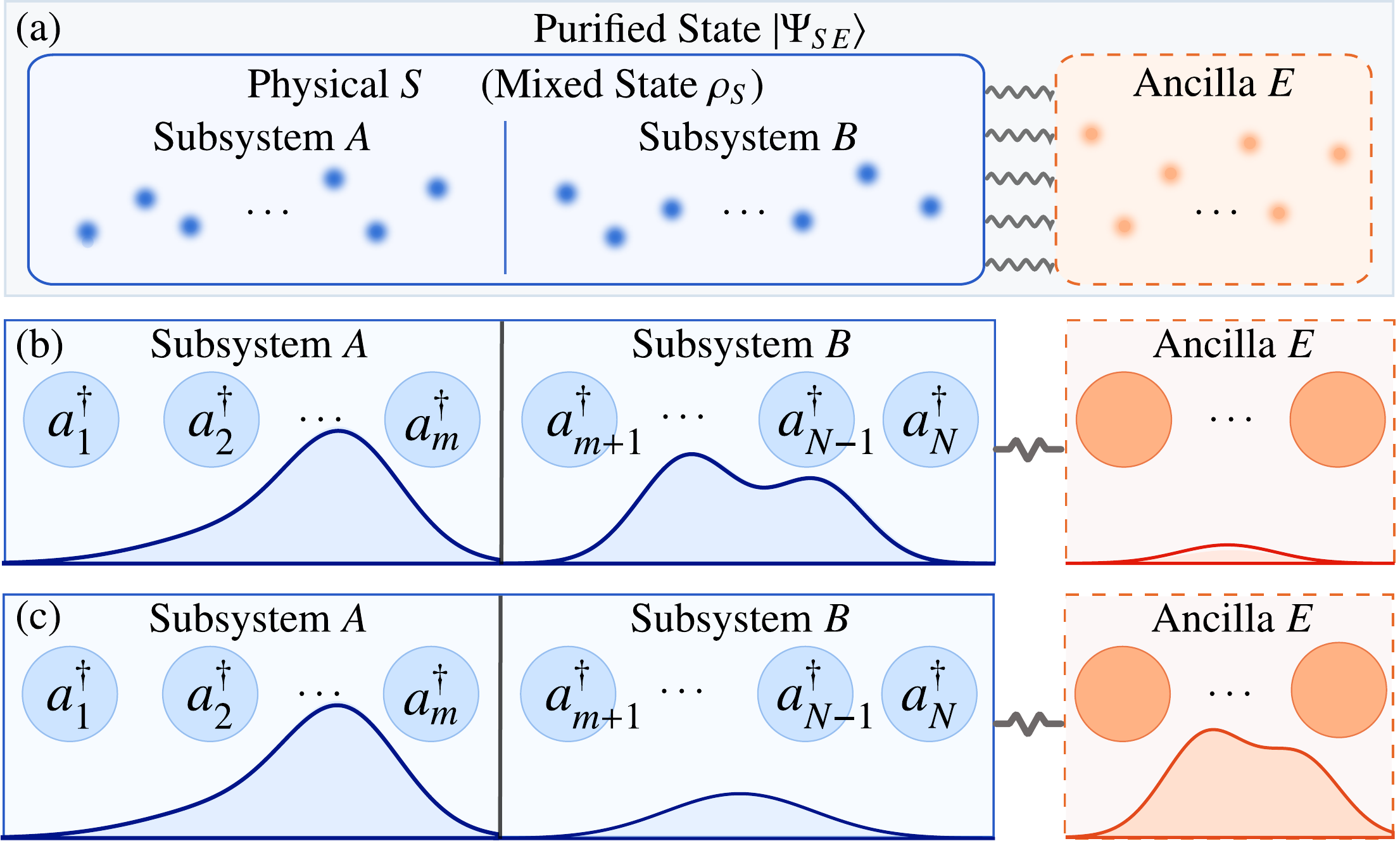}
\caption{
(a) Purification picture underlying the entanglement projected entropy. A mixed state $\rho_S$ on the physical bipartition $S=A\oplus B$ is embedded into a pure state $|\Psi_{SE}\rangle$ on the enlarged Hilbert space $A\oplus B\oplus E$, such that
$\rho_S=\Tr_E |\Psi_{SE}\rangle\langle\Psi_{SE}|$.
(b),(c) Schematic illustration of the spatial filtering mechanism in Gaussian states. Blue circles denote physical fermionic modes in $A$ and $B$, while orange circles denote ancilla modes in $E$. The curve drawn over subsystem $A$ shows the spatial wavefunction amplitude of a single-particle entanglement mode $\vec{\chi}_q$ on the sites in $A$, while the curve over $B\oplus E$ shows the amplitude distribution of its unique purification partner $\vec{\eta}_q$. The {\rm purification-independent} weight $w_q$ in Eq.~\eqref{eq:wq_physical_main} measures the fraction of the partner support residing in the physical subsystem $B$. 
(b) For $w_q\simeq 1$, the partner mode is predominantly supported in $B$, and this channel contributes significantly to the EPE in Eq.~\eqref{eq:PEE_definition}. 
(c) For $w_q\simeq 0$, the partner mode is predominantly supported in $E$, so this channel is filtered out.
}
    \label{fig:projected_entropy}
\end{figure}

To distinguish spatial entanglement across the bipartition $A\oplus B$ from mixed-state contributions, we embed $\rho_S$ into a globally pure Gaussian state $|\Psi_{SE}\rangle$ by introducing an ancilla environment $E$~\cite{Bravyi2004, NielsenChuang,Bennet2021}; see Fig.~\ref{fig:projected_entropy}(a). This purification provides a rigorous Gaussian canonical decomposition across the split $A|(BE)$, allowing us to resolve the entropy of $A$ into a set of \emph{entanglement channels}.

Consider the spectral decomposition of the local covariance block
$C_A \vec{\chi}_q = \lambda_q \vec{\chi}_q,$
where each wavefunction $\vec{\chi}_q$ represents a fermionic mode localized in subsystem $A$. In the purified Gaussian state, every such mode is paired, up to rotations within degenerate subspaces, with a normalized partner mode $\vec{\eta}_q$ supported in the complementary space $B\oplus E$~\cite{SM}. The entropy $s(\lambda_q)$ associated with this pair is the entropy contribution of the $q$-th channel.  The central question is how to assign it: if $\vec{\eta}_q$ is predominantly supported in $B$, the channel represents spatial entanglement across the physical cut; if it is predominantly supported in $E$, it is attributed to non-spatial mixed-state contributions.

This motivates the \emph{directional projection weight} $w_q = (\vec{\eta}_q)^\dagger P_B \vec{\eta}_q \in [0,1],$
which measures the fraction of the $q$-th partner mode projected into the physical subsystem $B$. As illustrated in Fig.~\ref{fig:projected_entropy}(b,c), channels with $w_q\simeq 1$ are predominantly purified by $B$ and therefore contribute strongly to spatial entanglement, whereas channels with $w_q\simeq 0$ are predominantly purified by the environment and are filtered out.

We therefore introduce the {\em EPE} by weighting each Schmidt channel in Eq.~\eqref{eq:vNE_gaussian}  according to its weight $w_q$ as
\begin{equation}
    \mathcal{E}_B(A) = \sum_q w_q s(\lambda_q).
    \label{eq:PEE_definition}
\end{equation} 
Remarkably, evaluating this spatial filter requires no explicit knowledge of the ancilla $E$; it can be extracted directly from the physical correlation matrix $C$. Since the spatial projection onto $B$ is governed by the cross-correlation block $C_{AB} = P_A C P_B$, the directional weight takes the simple form~\cite{SM}:
\begin{equation}
    w_q = \frac{1}{1-\lambda_q^2}(\vec{\chi}_q)^\dagger C_{AB}C_{BA} \vec{\chi}_q.
\label{eq:wq_physical_main}
\end{equation}
By substituting Eq.~\eqref{eq:wq_physical_main} back into the definition of the EPE and identifying $K_{AB}=C_{AB}C_{BA}$, we arrive at the central result of this work:
\begin{equation}
    \mathcal{E}_B(A)
    =
    \Tr_A\!\left[
        K_{AB}\,
        \frac{s(C_A)}{I_A-C_A^2}
    \right].
\label{eq:projected_entropy_trace_main}
\end{equation}
The trace is understood on the entangled subspace $|\lambda_q|<1$, since modes with $\lambda_q=\pm1$ have $s(\lambda_q)=0$ and do not contribute.
Although motivated by a purification, we emphasize that the final expression is purification independent: it involves only the physical covariance matrix $C$. 

Note that the EPE does not obey $A$-$B$ duality in a mixed state, i.e., $\mathcal{E}_B(A)\neq \mathcal{E}_A(B)$. This directionality is natural for a spatial filter: $\mathcal E_B(A)$ asks how much of the entropy of $A$ is purified by $B$. Nevertheless, it has the correct limiting behavior: In the pure-state limit, global purity enforces the exact block identity $K_{AB}=I_A-C_A^2$~\cite{SM}, so that $\mathcal{E}_B(A)$ reduces to the standard bipartite entanglement entropy. Conversely, in the infinite-temperature limit, we have $C_{AB}=0$, implying $K_{AB}\to 0$ and hence $\mathcal{E}_B(A)=0$, as expected in the absence of spatial coherence across the cut. 

An important structural property is regional additivity. For a multipartite decomposition into disjoint physical regions, $S=A\oplus B\oplus C$, one finds~\cite{SM}
\begin{equation}
    \mathcal E_{B\oplus C}(A)=\mathcal E_B(A)+\mathcal E_C(A).\label{eq:region_add}
\end{equation}
This additivity indicates that the complement-purified part of the entropy can be decomposed according to disjoint physical regions.

The key role of the EPE is therefore to remove extensive mixed-state backgrounds while retaining boundary-sensitive quantum correlations. We now demonstrate this capability in two critical settings: a finite-temperature 1D free-fermion chain, and a 2D $\pi$-flux model. As a complementary gapped benchmark, we show in the Supplemental Material~\cite{SM} that $\mathcal{E}_B(A)$ also resolves the finite-temperature remnant of the $\ln 2$ boundary contribution~\cite{Ryu2006, Fidkowski2010} in the topological SSH chain~\cite{Su1979}.

{\em Universal CFT scaling in 1D gapless chain.---}
We first test the EPE in a critical gapless setting, where the ordinary finite-temperature vNE is notoriously dominated by an extensive thermal volume law. A natural benchmark is the 1D spinless fermion chain defined by the Hamiltonian
\begin{equation}
    H=t\sum_{j=1}^{L}\left(
     a_{j}^\dagger a_{j+1} + 
     a_{j+1}^\dagger a_{j}
    \right).
\label{eq:chain_Hamiltonian}
\end{equation}
At half filling, Eq.~\eqref{eq:chain_Hamiltonian} has a Fermi velocity 
$v_F=2t$ at $k_F=\pi/2$.
This system is effectively described by a $(1+1)$-dimensional free-fermion CFT~\cite{DiFrancesco1997,Giamarchi2003}, whose zero-temperature entanglement entropy obeys the universal logarithmic scaling governed by the central charge $c=1$~\cite{Holzhey1994,Calabrese2004,Vidal2003,Korepin2004}. This makes it an ideal platform to examine whether the EPE can recover the CFT structure from a thermal background. 

\begin{figure}[!t]
    \centering
    \includegraphics[width=1.\linewidth]{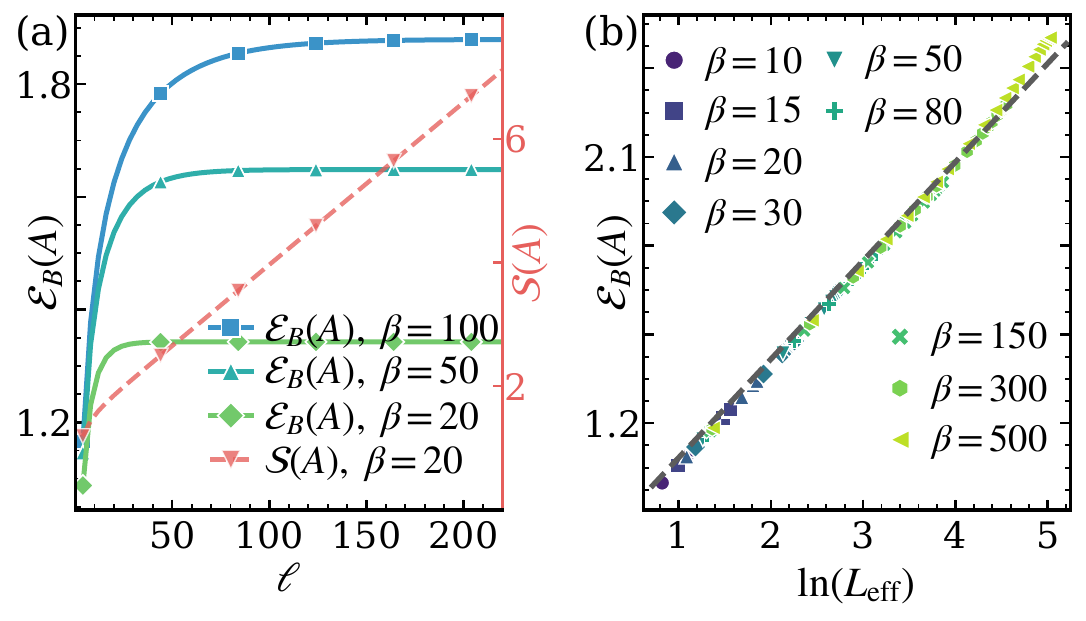} 
\caption{
    Numerical verification of the universal CFT scaling in an infinite 1D free-fermion chain at finite temperature. 
    (a) Projected entropy $\mathcal{E}_B(A)$ (left axis, solid lines) and standard von Neumann entropy $\mathcal{S}(A)$ (right axis, red dashed line) as a function of the subsystem size $\ell$. While $\mathcal{S}(A)$ exhibits an extensive thermal volume law (linear in $\ell$), $\mathcal{E}_B(A)$ cleanly removes this thermal background and saturates to a plateau for $\ell \gg v_F\beta$. 
    (b) Data collapse of $\mathcal{E}_B(A)$ against the effective conformal length $\ln(L_{\rm eff})$, with $L_{\rm eff}$ defined in Eq.~\eqref{eq:Leff}. The dashed gray line indicates the analytical CFT prediction with a slope of $c/3$ ($c=1$).
}
    \label{fig:CFT}
\end{figure}

For analytical control, we work in the continuum limit and consider a half-filled free fermion chain on the infinite real line at inverse temperature $\beta$. We take the subsystem as $A=(0,\ell)$ and complement $B=(-\infty,0)\oplus(\ell,\infty)$. In this continuum formulation, the discrete correlation matrices~\cite{Peschel2003,PeschelEisler2009} are naturally replaced by spatial integral operators~\cite{CasiniHuerta2009}. 
The low-energy theory of Eq.~\eqref{eq:chain_Hamiltonian} consists of right- and left-moving chiral branches. For the universal part, we work branch by branch with the equal-time Cauchy kernel $C^{(0)}(y_1,y_2)=\frac{1}{2\pi i}\frac{1}{y_1-y_2} $ ($y_1\neq y_2$), regulated by a short-distance cutoff $a$~\cite{Giamarchi2003}; the two branches are included through the total central charge \(c=1\).

At finite temperature, the physical cylinder geometry can be mapped to the complex plane via the standard conformal transformation~\cite{Calabrese2004} 
\begin{equation}
y(x)=\frac{v_F\beta}{2\pi}\left(1-e^{-2\pi x/v_F\beta}\right).    \label{eq:cft_mapping}
\end{equation}
As a primary field of conformal dimension $1/2$~\cite{DiFrancesco1997}, the finite-temperature correlation kernel transforms as
\begin{equation}
    C^\beta(x_1,x_2)
    =
    \sqrt{y'(x_1)y'(x_2)}
    \,C^{(0)}\left(y(x_1),y(x_2)\right),
\end{equation}
where $y'(x)=dy/dx.$
The continuous spatial filter entering the EPE is correspondingly given by $K_{AB}^\beta(x_1,x_2)=\int_B dz\, C^\beta(x_1,z)C^\beta(z,x_2)$. 

Evaluating the continuous trace in Eq.~\eqref{eq:projected_entropy_trace_main} directly is difficult.  
Nevertheless, the conformal map gives a simple geometric
interpretation of the EPE. Using Eq.~\eqref{eq:cft_mapping}, the physical line is mapped to the image domain $(-\infty,v_F\beta/2\pi)$ and the interval $A=(0,\ell)$ becomes
\begin{equation}
    A_{\rm eff}=(0,L_{\rm eff}),
    \qquad
    L_{\rm eff}=\frac{v_F\beta}{2\pi}\left(1-e^{-2\pi \ell/v_F\beta}\right).\label{eq:Leff}
\end{equation}
Meanwhile, the physical complement is mapped only to
\begin{equation}
    B_{\rm eff}=(-\infty, ~0)\oplus(L_{\rm eff}, ~v_F\beta/2\pi),
\end{equation}
rather than to the full complement of $A_{\rm eff}$ on the vacuum line.
The remaining region $(v_F\beta/2\pi, ~\infty)$ plays the role of the thermal purification sector in the image representation. Thus the projected entropy should be viewed as the part of the vacuum entanglement contour of $A_{\rm eff}$ supported on the physical image complement $B_{\rm eff}$, while the missing part corresponds to thermal mixing~\cite{SM}.

Using the CFT complement-side contour for an interval, the contribution given by the physical complement $B_{\rm eff}$ leads to~\cite{SM}
\begin{equation}
    \mathcal{E}_B(\ell,\beta)
    \simeq
    \frac{c}{3}
    \ln\!\left[
        \frac{v_F\beta}{2\pi a}
        \left(1-e^{-2\pi\ell/v_F\beta}\right)
    \right]
    +\mathrm{const},\quad c=1.
\end{equation}
Note that it successfully reproduces the pure-state logarithm $\mathcal{E}_B \simeq \frac{c}{3}\ln(\ell/a)$ for short intervals ($\ell\ll v_F\beta$), and saturates to a finite plateau $\mathcal{E}_B \simeq \frac{c}{3}\ln(\frac{v_F\beta}{2\pi a})$ in the limit $\ell\gg v_F\beta$. 
Equivalently, comparing with the standard finite-temperature CFT entropy~\cite{Calabrese2004}
$S_\beta(A)=\frac{c}{3}\ln\left[
\frac{v_F\beta}{\pi a}\sinh\frac{\pi\ell}{v_F\beta}
\right]+\mathrm{const},$
one obtains
\begin{equation}
\mathcal E_B(A) = S_\beta(A) - \frac{\pi c}{3v_F\beta}\ell +\mathrm{const}.
\end{equation}
This result explicitly demonstrates that the thermal volume law is removed while the underlying CFT scaling is retained. 
Related universal thermal corrections were analyzed in Ref.~\cite{CardyHerzog2014}.

To verify this analytical continuum prediction, we numerically evaluate the EPE for the lattice model in Eq.~\eqref{eq:chain_Hamiltonian}. As shown in Fig.~\ref{fig:CFT}(a), the standard finite-temperature vNE $\mathcal{S}(A)$ is severely dominated by an extensive thermal volume law, growing linearly with the subsystem size $\ell$. In stark contrast, the EPE $\mathcal{E}_B(A)$ removes this leading extensive thermal background. It saturates to a finite and temperature-dependent plateau for $\ell \gg v_F\beta$.

Furthermore, to quantitatively benchmark the universal CFT scaling, we plot $\mathcal{E}_B(A)$ against the effective conformal length $\ln(L_{\rm eff})$ in Fig.~\ref{fig:CFT}(b). Remarkably, across a wide range of inverse temperatures $\beta$, all data points collapse onto a single universal straight line. The slope of this line accurately matches the $c/3$ ($c=1$) prefactor expected from the underlying CFT.

\begin{figure}
    \centering
    \includegraphics[width=1\linewidth]{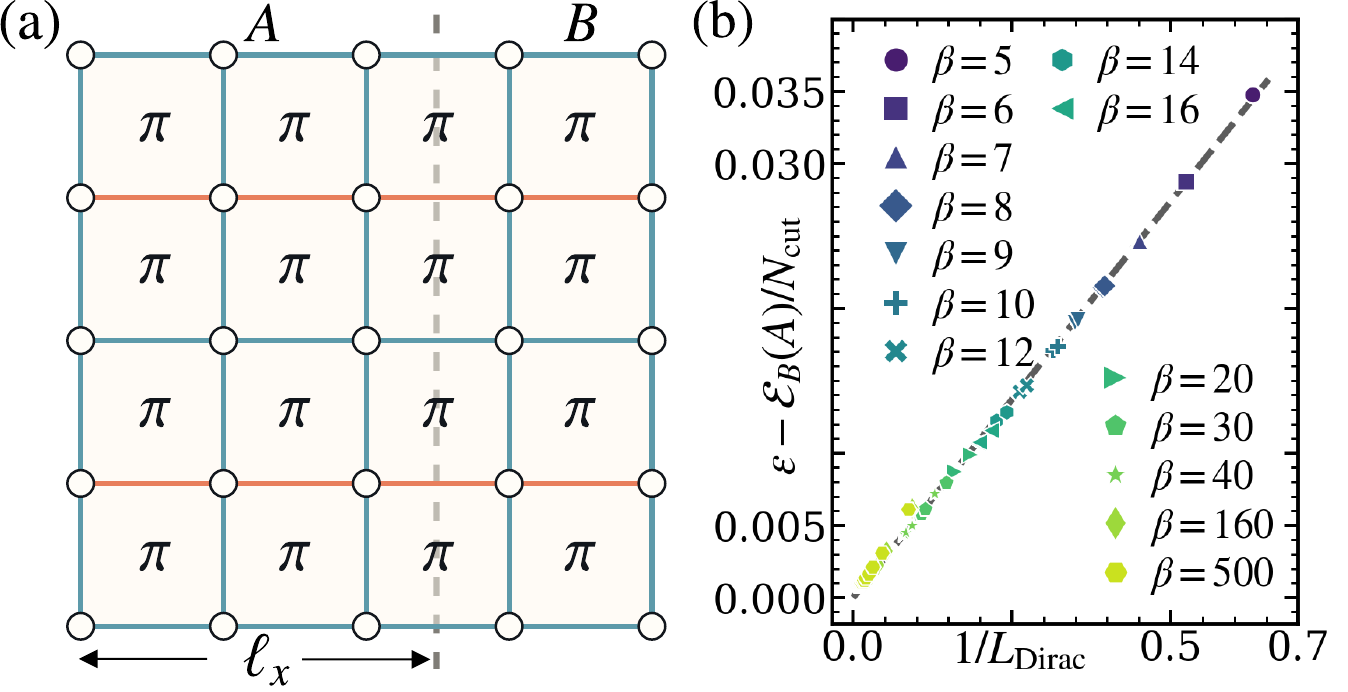}
\caption{(a) Half-filled $\pi$-flux model on a square lattice. Blue and red bonds denote hopping amplitudes $-t$ and $t$, respectively,  yielding a $\pi$ flux per plaquette. The dashed line with length $L_y$ marks the bipartition into $A$ and $B$, with $A$ a vertical strip of width $\ell_x$.  
(b) Scaling collapse of the EPE density correction
$\left[\epsilon N_{\rm cut}-\mathcal{E}_B(A)\right]/N_{\rm cut}$
versus $1/L_{\rm Dirac}$, with $N_{\rm cut}=2L_y$ and $L_{\rm Dirac}$ defined in Eq.~\eqref{eq:Leff2d}. The dashed line indicates the linear dependence in Eq.~\eqref{eq:piflux_scaling_law}.
}\label{fig:2dcft}
\end{figure}

{\em Scaling of 2+1D Dirac fermions.---}
The success of EPE in 1D chain relies on the exact conformal mapping of $(1+1)$D CFT. To examine whether $\mathcal{E}_B(A)$ can distill universal infrared signatures in higher dimensions, we next consider a two-dimensional gapless system: the half-filled $\pi$-flux model on the square lattice~\cite{Affleck1988}. 
This model serves as a paradigmatic lattice realization of massless Dirac fermions in $2+1$ dimensions. As illustrated in Fig.~\ref{fig:2dcft}(a), the Hamiltonian is given by
\begin{equation}
    H = -t\sum_{i_x,i_y} \left[ (-1)^{i_y} c^\dagger_{i_x+1,i_y}c_{i_x,i_y} + c^\dagger_{i_x,i_y+1}c_{i_x,i_y} + \mathrm{h.c.} \right],
    \label{eq:piflux_H}
\end{equation}
where the phase factor $(-1)^{i_y}$ ensures a $\pi$ magnetic flux through each plaquette~\cite{Affleck1988}. The dispersion relation of $\pi$-flux state is $E_\pm(\mathbf{k}) = \pm 2t\sqrt{\cos^2 k_x + \cos^2 k_y}$, which exhibits isolated Dirac nodes at $\mathbf{k}^* = (\pm \pi/2, \pm \pi/2)$. Near these nodes, the low-energy physics is governed by a $(2+1)$D CFT~\cite{Chen2015, Chen2017} with a Fermi velocity $v_F=2t$. 

We study the system numerically on a torus of size $L_x\times L_y$ with periodic boundary conditions along $x$ and anti-periodic boundary conditions along $y$. The latter choice prevents the allowed momenta from passing directly through the Dirac nodes~\cite{Chen2017,Jin2025}, thereby avoiding accidental ground-state degeneracies. We bipartition the torus into a vertical strip subsystem $A$ of width $\ell_x$ and its complement $B$; see Fig.~\ref{fig:2dcft}(a).

At zero temperature, the entanglement entropy of a torus strip in a $(2+1)$D scale-invariant theory takes the form $\mathcal{S}(A)\big|_{T=0}
    =
    \epsilon\, N_{\rm cut}
    +
    S_{\rm strip}
    +
    \cdots ,$
where $N_{\rm cut}=2L_y$ is the total length of the two entangling cuts and $\epsilon$ is the nonuniversal area-law coefficient per unit cut length~\cite{Chen2015,Chen2017,Casini2007}.
The first term is the nonuniversal area-law contribution~\cite{Srednicki1993,EisertCramerPlenio2010}, while $S_{\rm strip}$ is a universal shape-dependent subleading term~\cite{Chen2015,Chen2017}.
Motivated by this rich structure, we ask how these boundary-controlled correlations manifest at $T>0$ when the entanglement entropy is replaced by $\mathcal{E}_B(A)$.

The finite-temperature scaling can be understood from dimensional reduction. Due to the translation symmetry along $y$, the single-particle Hamiltonian and the Gaussian correlation matrix decompose into transverse-momentum sectors. This indicates that the projected entropy also decomposes into transverse-momentum sectors as $\mathcal{E}^{(2D)}_B(A)=\sum_{k_y}\mathcal{E}^{(1D)}_B(k_y, \ell_x)$. For generic $k_y$, the deviation $q_y$ from a Dirac node acts as an effective mass $m \propto v_F|q_y|$~\cite{Casini2005}, making the $(1+1)$D mode gapped at $T=0$. The contributions of these gapped modes merely renormalize the leading area-law term $\epsilon N_{\rm cut}$. The nontrivial finite-temperature dependence arises solely from the sectors in the vicinity of the Dirac nodes. For these modes, both the temperature $T$ and the small effective mass $m$ act as infrared regulators for the singular $(1+1)$D contribution. Consequently, their contribution is governed by the same infrared physics as in $(1+1)$D and can be described by a scaling function of $\ell_x$ and $v_F\beta$.

Motivated by our $(1+1)$D finite-temperature conformal mapping~\cite{Calabrese2004}, we introduce the infrared length
\begin{equation}
    L_{\rm Dirac}(\ell_x,\beta)
    =
    \frac{v_F\beta}{2\pi}
    \left(1-e^{-2\pi \ell_x/v_F\beta}\right),
    \label{eq:Leff2d}
\end{equation}
which interpolates between the spatial width $\ell_x$ at low temperature and the thermal scale $v_F\beta/2\pi$ at high temperature. 
The singular contribution of each transverse sector follows the $(1+1)$D logic: in the massless limit it scales as $\sim \ln L_{\rm Dirac}$. 
For a momentum $q_y$ away from a Dirac node, the effective mass $m\simeq v_F|q_y|$ cuts off this logarithm at the mass length $|q_y|^{-1}$. 
After subtracting the contribution of massive sectors, the remaining finite-$L_{\rm Dirac}$ correction is controlled by the scaling variable $|q_y|L_{\rm Dirac}$ and hence by an infrared window $|q_y|\lesssim L_{\rm Dirac}^{-1}$. 
The number of such modes scales as $L_y/L_{\rm Dirac}$, giving a correction per boundary length proportional to $1/L_{\rm Dirac}$.

We verify this scaling analysis numerically in Fig.~\ref{fig:2dcft}(b).  The correction of the EPE density relative to the ground-state area law is computed on a torus with $L_x=600$ and $L_y=300$.
Remarkably, data for different inverse temperatures $\beta$ and strip widths $\ell_x$ collapse onto a straight line, consistent with
\begin{equation}
    \epsilon -\mathcal{E}_B(A)/N_{\rm cut}
    \propto L_{\rm Dirac}^{-1}(\ell_x,\beta).
    \label{eq:piflux_scaling_law}
\end{equation}
Here $\epsilon$ is fixed by a zero-temperature covariance-matrix calculation on the same finite-size lattice.
This collapse supports the dimensional-reduction picture: after the nonuniversal area-law density is removed, the remaining EPE correction is controlled by the Dirac infrared scale rather than by $\ell_x$ or $\beta$ separately.
This provides evidence that $\mathcal{E}_B(A)$ filters out the extensive thermal background and exposes the boundary-controlled infrared correlations of the Dirac state.

{\em Discussion.---}
In summary, we have introduced EPE as a Gaussian spatial filter for mixed-state subsystem entropy, with a closed-form expression in terms of the physical covariance matrix.
In a finite-temperature 1D critical free-fermion chain, it removes the extensive thermal background and recovers the universal CFT logarithmic scaling.
In the 2D half-filled $\pi$-flux model, it reveals a parameter-free scaling collapse governed by the Dirac infrared length.

An important open problem is to extend this spatial-filtering idea to interacting many-body mixed states. The Gaussian construction suggests a possible guiding principle: in an interacting many-body system, the relevant projected weights may need to be assigned to operators connecting
different eigenstates of $\rho_A$, rather than directly to the eigenvalues of $\rho_A$. These operators have natural partner
operators in a purification, reminiscent of modular-theoretic mirror
constructions~\cite{Haag2012}.
The guiding goal is to quantify how much of the entropy of $A$ is associated with correlations to its physical complement, rather than with local thermal mixing. 
This perspective may be useful for extracting boundary-sensitive correlations in finite-temperature quantum critical systems, Gaussian open-system steady states, and numerical studies of mixed-state many-body systems.

\textit{Acknowledgement.---}
We acknowledge Hong-Hao Tu and Yi Zhou for useful discussions. 
H.-K. J. acknowledges support from the National Natural Science Foundation of China (NSFC-12504180) and the start-up funding from ShanghaiTech University.

\bibliography{main}

\clearpage

\appendix

\clearpage

\begin{widetext}
\begin{center}

{\large\bf Supplementary Material for}
{\large\bf ``Extracting Universal Entanglement Scaling from Mixed Fermionic Gaussian States via Entanglement Projected Entropy''}
\vspace{0.3cm}
\\
Jia-Wen Tao$^1$ and Hui-Ke Jin$^{1,*}$
\\
{\em $^1$School of Physical Science and Technology, ShanghaiTech University, Shanghai 201210, China}
\vspace{0.4cm}
\end{center}
\end{widetext}

In this Supplemental Material, we provide technical derivations and additional results that complement the main text. 
We first present the Gaussian formulation of the entanglement projected entropy (EPE), including the purification construction, the partner-mode projection weight, and its closed-form expression in terms of the physical covariance matrix. 
We then discuss the additivity of the EPE for decoupled Gaussian sectors. 
As a complementary benchmark, we study the SSH chain and show that the EPE resolves the finite-temperature remnant of the topological $\ln 2$ boundary contribution. 
Finally, we derive the finite-temperature EPE in the one-dimensional critical free-fermion chain using a conformal-mapping argument.

\section{Gaussian formulation of entanglement projected entropy}

In this section, we provide the microscopic operator construction of the fermionic Gaussian states and rigorously derive the minimal purification formalism used in the main text. 

\subsection{Schmidt decomposition and entanglement in pure Gaussian states}

We consider $N$ spinless fermionic modes with creation operators $a_j^\dagger$ ($j=1,\dots,N$). A pure Gaussian state with $Q$ occupied single-particle orbitals can be written as
\begin{equation}
    |\phi\rangle=\prod_{q=1}^{Q} d_q^\dagger |0\rangle,
\end{equation}
where $d_q^\dagger=\sum_{j=1}^{N} a_j^\dagger V_{jq},$ and $V$ is an $N\times N$ unitary matrix.

The state is fully characterized by the centered covariance matrix
\begin{equation}
    C_{jl}=2\langle \phi|a_l^\dagger a_j|\phi\rangle-\delta_{jl}~\Longrightarrow ~
    C=V
    \begin{pmatrix}
        \mathbb{I}_{Q} & 0\\
        0 & -\mathbb{I}_{N-Q}
    \end{pmatrix}
    V^\dagger.
\end{equation}
It is easy to verify that $C^\dagger=C$ and
\begin{equation}
    C^2=\mathbb{I}_N.
\end{equation}

We now bipartition the system into $A\oplus B$, where $A$ contains $m$ modes and $B$ contains $\bar m=N-m$ modes. Let $P_A$ and $P_B$ be the projectors onto the corresponding single-particle subspaces, with
\begin{equation}
    P_A+P_B=\mathbb{I}, \qquad P_A P_B=0.
\end{equation}
The covariance matrix then decomposes as
\begin{equation*}
\begin{split}    
    &C_A=P_A C P_A,\quad
    C_B=P_B C P_B,\quad \\
    &C_{AB}=P_A C P_B,\quad
    C_{BA}=P_B C P_A,
\end{split}
\end{equation*}
or equivalently,
\begin{equation}
    C=
    \begin{pmatrix}
        C_A & C_{AB}\\
        C_{BA} & C_B
    \end{pmatrix}.
\end{equation}
Note that $C$ is Hermitian, i.e., $C_{AB}=C_{BA}^\dagger.$
Since the global state is pure, $C^2=\mathbb{I}$ implies that
\begin{align}
    C_A^2 + C_{AB}C_{BA} &= \mathbb{I}_A, \label{eq:pure_block1}\\
    C_B^2 + C_{BA}C_{AB} &= \mathbb{I}_B, \label{eq:pure_block2}\\
    C_A C_{AB} + C_{AB} C_B &= 0. \label{eq:pure_block3}
\end{align}
Let $\vec{\chi}_q$ be one of the eigenmodes of $C_A$
\begin{equation}
    C_A \vec{\chi}_q=\lambda_q \vec{\chi}_q,
    \qquad \lambda_q\in[-1,1],
\end{equation}
which is normalized as $(\vec{\chi}_q)^\dagger \vec{\chi}_q = 1$. Modes with $\lambda_q=\pm1$ are locally pure and do not contribute to bipartite entanglement. For each nontrivial mode with $|\lambda_q|<1$, Eqs.~\eqref{eq:pure_block1}--\eqref{eq:pure_block3} imply that it is paired with a unique mode in $B$. Defining
\begin{equation}
    \vec{\eta}_q
    =
    \frac{1}{\sqrt{1-\lambda_q^2}}\,C_{BA}\vec{\chi}_q,\label{eq:right_mode}
\end{equation}
one finds that $\vec{\eta}_q$ is normalized and satisfies
\begin{equation}
    C_B \vec{\eta}_q=-\lambda_q \vec{\eta}_q.
\end{equation}
Thus each entangled channel forms a two-mode pair across the cut $A|B$.

Denoting $(\vec{\chi}_q)_j$ as the $j$-th entry of vector $\vec{\chi}_q$, we define the canonical fermion operators as 
\begin{equation}
    \alpha_q^\dagger=\sum_{j\in A} a_j^\dagger (\vec{\chi}_q)_j,
    \qquad
    \beta_q^\dagger=\sum_{j\in B} a_j^\dagger (\vec{\eta}_q)_j.
\end{equation}
The pure Gaussian state is then factorized into independent Schmidt pairs,
\begin{equation}
\begin{split}
    |\phi\rangle
    &=
    \left(
        \prod_{r\in\mathbb{O}_A}
        \alpha_r^\dagger
    \right)
    \left(
        \prod_{s\in\mathbb{O}_B}
        \beta_s^\dagger
    \right)
    \left[
        \prod_{q\in\mathbb{E}}
        \left(
            \sqrt{\frac{1+\lambda_q}{2}}\,
            \alpha_q^\dagger
            +
            \sqrt{\frac{1-\lambda_q}{2}}\,
            \beta_q^\dagger
        \right)
    \right]
    |0\rangle ,
\label{eq:paired_state}
\end{split}
\end{equation}
where $\mathbb{E}$ denotes the set of modes with $|\lambda_q|<1$, $\mathbb{O}_A$ denotes the occupied single-particle
orbitals supported entirely in $A$, corresponding to eigenmodes of
$C_A$ with eigenvalue $+1$; and $\mathbb{O}_B$ denotes the occupied
single-particle orbitals supported entirely in $B$, corresponding to
eigenmodes of $C_B$ with eigenvalue $+1$. Eq.~\eqref{eq:paired_state} indicates that the total correlation weight emitted from mode $\vec{\chi}_q$ into the rest of the universe ($B$) is exactly $1-\lambda_q^2$, by noting that $\alpha_q^\dagger$ and $\beta_q^\dagger$ are normalized.

The bipartite entanglement entropy is then the sum of independent single-mode contributions,
\begin{equation}
    \mathcal{S}(A)=\sum_{q\in\mathbb{E}} s(\lambda_q)=\Tr_A[s(C_A)].
\end{equation}
with
\begin{equation}
    s(\lambda)=
    -\frac{1+\lambda}{2}\ln\frac{1+\lambda}{2}
    -\frac{1-\lambda}{2}\ln\frac{1-\lambda}{2}.
\end{equation}

\subsection{Minimal Gaussian purification and the projection weight $w_q$}

We now turn to a mixed Gaussian state defined on the full physical system $S$. For mixed state, we have ${C}^2\neq{}\mathbb{I}$. 
We purify the state by introducing an ancilla fermionic environment $E$ with $N$ ancilla modes $\gamma_j^\dagger$. The pure Gaussian state on the enlarged space $S\oplus E$ has the block covariance matrix
\begin{equation}
    \tilde C=
    \begin{pmatrix}
        C & C_{SE}\\
        C_{ES} & C_E
    \end{pmatrix},
\end{equation}
which strictly satisfies $\tilde C^2=\mathbb{I}_{S\oplus E}$. Note that the form of purification is not unique and later we will show that its explicit form does not affect our results. 

We bipartition the physical system into $S=A\oplus B$, leaving the purified $\tilde{C}$ a pure tripartite Gaussian state on $A\oplus B\oplus E$:
\begin{equation}
    \tilde C=
    \begin{pmatrix}
        C_A & C_{AB} & C_{AE}\\
        C_{BA} & C_B & C_{BE}\\
        C_{EA} & C_{EB} & C_E
    \end{pmatrix}.
\end{equation}
where the top-left $2\times 2$ block is the physical mixed state $C$. 
Because the extended tripartite state is globally pure, it strictly satisfies $\tilde{C}^2 = \mathbb{I}_{A \oplus B \oplus E}$. Evaluating the $A$-$A$ diagonal block of this matrix identity yields:
\begin{equation}
    C_A^2 + C_{AB}C_{BA} + C_{AE}C_{EA} = \mathbb{I}_A.\label{eq:purity_A}
\end{equation}

For the $q$-th left canonical mode $\vec{\chi}_q$ defined by $C_A \vec{\chi}_q = \lambda_q \vec{\chi}_q$, Eq.~(\ref{eq:purity_A}) gives:
\begin{equation}
    \left(C_{AB}C_{BA} + C_{AE}C_{EA}\right) \vec{\chi}_q= (1 - \lambda_q^2)\vec{\chi}_q.
\end{equation}
We focus on the case of $\lambda_q\neq{}\pm{}1$. We define the block matrices connecting $A$ and its full complement $\bar{A} = B \oplus E$ as:
\begin{equation}
    C_{A\bar{A}} = \begin{pmatrix} C_{AB} & C_{AE} \end{pmatrix}, \quad
    C_{\bar{A}A} = \begin{pmatrix} C_{BA} \\ C_{EA} \end{pmatrix}.
\end{equation}
As indicated by Eq.~\eqref{eq:right_mode}, the normalized right-mode in the space $B\oplus E$ is:
\begin{equation}
    \vec{\eta}_q = \frac{1}{\sqrt{1 - \lambda_q^2}} C_{\bar{A}A} \vec{\chi}_q.
\end{equation}
The physical weight $w_q$ is the expectation value of the spatial projector $P_B$ on this right-mode:
\begin{equation}
\begin{split}
w_q & = (\vec{\eta}_q)^\dagger P_B \vec{\eta}_q \\
&= \frac{(\vec{\chi}_q)^\dagger C_{A\bar{A}} P_B C_{\bar{A}A} \vec{\chi}_q}{1 - \lambda_q^2}.\label{eq:wq1}
\end{split}
\end{equation}
In the enlarged space $B \oplus E$, the projector onto $B$ takes the block form $P_B = \begin{pmatrix} \mathbb{I}_B & 0 \\ 0 & 0_E \end{pmatrix}$. Consequently, $C_{A\bar{A}} P_B C_{\bar{A}A} = C_{AB}C_{BA}$, and Eq.~\eqref{eq:wq1} can be simplified as
\begin{equation}
    w_q = \frac{(\vec{\chi}_q)^\dagger {C}_{AB} {C}_{BA} \vec{\chi}_q}{1 - \lambda_q^2}.\label{eq:wq2}
\end{equation}

Finally, we note the edge case of completely pure modes with $\lambda_q = \pm 1$. In this case, the variance $1 - \lambda_q^2$ vanishes, and Eq.~(\ref{eq:purity_A}) dictates that the cross-correlation also strictly vanishes, e.g., $C_{\bar{A}A} \vec{\chi}_q = 0$. Such modes are entirely decoupled from the rest of the system and carry zero entanglement entropy, e.g., $s(\pm 1) = 0$. Consequently, they do not contribute to the total EPE $\mathcal{E}_B(A)$ and are simply excluded from the definition of $w_q$ and the subsequent trace.

\section{Additivity of the EPE}
\label{app:additivity_projected_entropy}

In this appendix we prove the additivity property
\begin{equation}
    \mathcal E_{B\oplus C}(A)
    =
    \mathcal E_B(A)+\mathcal E_C(A),
\end{equation}
for two disjoint regions \(B\) and \(C\) in the complement of \(A\).

For any region \(R\) belonging to the complement of $A$, define
\begin{equation}
    \mathcal E_R(A)
    =
    \Tr_A
    \left[
        K_{AR} W_A
    \right],
\end{equation}
where
\begin{equation}
    K_{AR}
    =
    C_{AR}C_{RA},
\end{equation}
and
\begin{equation}
    W_A
    =
    \frac{s(C_A)}{I_A-C_A^2}.
\end{equation}
The important point is that \(W_A\) depends only on the region \(A\), not on
the choice of \(R\).

Now take
\begin{equation}
    R=B\oplus C,
    \qquad
    B\cap C=\varnothing .
\end{equation}
In block-matrix form,
\begin{equation}
    C_{A,B\oplus C}
    =
    \begin{pmatrix}
        C_{AB} & C_{AC}
    \end{pmatrix},
\end{equation}
and
\begin{equation}
    C_{B\oplus C,A}
    =
    \begin{pmatrix}
        C_{BA}\\
        C_{CA}
    \end{pmatrix}.
\end{equation}
Therefore
\begin{align}
    K_{A,B\oplus C}
    &=
    C_{A,B\oplus C}C_{B\oplus C,A}
    \nonumber\\
    &=
    \begin{pmatrix}
        C_{AB} & C_{AC}
    \end{pmatrix}
    \begin{pmatrix}
        C_{BA}\\
        C_{CA}
    \end{pmatrix}
    \nonumber\\
    &=
    K_{AB}+K_{AC}.
\end{align}
Hence
\begin{align}
    \mathcal E_{B\oplus C}(A)
    &=
    \Tr_A
    \left[
        K_{A,B\oplus C}W_A
    \right]
    \nonumber\\
    &=
    \mathcal E_B(A)+\mathcal E_C(A).
\end{align}
This proves the additivity of the projected entropy over disjoint complement
regions.

\section{Boundary entanglement in SSH model}

We next turn to a gapped lattice model, where the key question is whether the EPE can retain the characteristic boundary contribution of a topological phase in the presence of thermal mixing. As a minimal setting, we study the SSH chain~\cite{Su1979},
\begin{equation}
    H=\sum_{j=1}^{L}\left(
    t_1\, a_{j,1}^\dagger a_{j,2}
    +t_2\, a_{j,2}^\dagger a_{j+1,1}
    +\mathrm{h.c.}
    \right),
\label{eq:SSH_Hamiltonian}
\end{equation}
where $a_{j,s}^\dagger$ creates a spinless fermion in unit cell $j$ on sublattice $s=1,2$. 
At half filling, the model has chiral symmetry and realizes two distinct zero-temperature phases: a trivial dimerized phase for $t_1>t_2$ and a symmetry-protected topological phase for $t_1<t_2$~\cite{Ryu2006}

\begin{figure}
    \centering
    \includegraphics[width=1\linewidth]{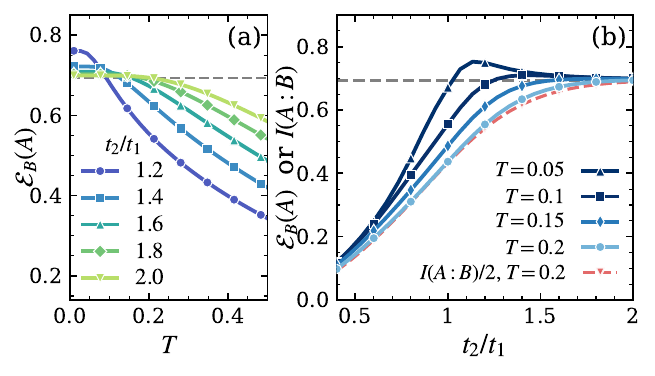}
\caption{
    Projected entropy in the SSH chain for a half-chain bipartition.
    (a) Temperature dependence of $\mathcal{E}_B(A)$ for several values of $t_2/t_1$ in the topological regime. The dashed line marks $\ln 2$.
    (b) Projected entropy $\mathcal{E}_B(A)$ as a function of $t_2/t_1$ for several fixed temperatures. For comparison, half of the mutual information, $I/2$, at $T=0.2$ is also shown as a dashed line, demonstrating its close agreement with $\mathcal{E}_B(A)$ at $T=0.2$.
}
    \label{fig:SSH}
\end{figure}

Under open boundary conditions, the zero-temperature topological phase is distinguished by a universal boundary contribution of $\ln 2$ to the entanglement entropy~\cite{Ryu2006, Fidkowski2010}, a feature that is strictly absent in the trivial phase. However, extending this topological diagnostic to finite temperatures is highly nontrivial: the standard vNE becomes dominated by the extensive thermal volume law, which completely obscures the sub-extensive $\ln 2$ signature. Remarkably, as we show below, the EPE filters out this thermal background and precisely recovers the robust $\ln 2$ topological plateau.

We evaluate the half-chain EPE $\mathcal{E}_B(A)$ in Eq.~\eqref{eq:projected_entropy_trace_main} for an SSH chain with length $L=120$. As shown in Fig.~\ref{fig:SSH}(a), $\mathcal{E}_B(A)$ recovers the quantized $\ln 2$ boundary signature at low temperatures. As $T$ increases, thermal fluctuations smoothly erode this topological entanglement, with the decay strongly suppressed deeper in the topological phase where the bulk gap is larger. The distilling power of $\mathcal{E}_B(A)$ becomes starkly evident when tuning across the phase diagram; see Fig.~\ref{fig:SSH}(b).

To benchmark the EPE against a standard measure of correlations in mixed states, we compare it with the mutual information (MI)~\cite{Wolf2008,Iaconis2013},
\begin{equation}
I=\mathcal{S}(A)+\mathcal{S}(B)-\mathcal{S}(A\oplus B),\label{eq:MI}   
\end{equation}
which can also suppress thermal volume-law contributions.
For a bipartition of a pure state, it satisfies $I(A\!:\!B)=2S(A)$ and therefore yields $\approx{}2\ln 2$ for the bipartition in the topological phase at zero temperature.
We find that, throughout the SSH parameter regime considered here, the EPE $\mathcal{E}_B(A)$ closely tracks one-half of the mutual information, $I/2$. This shows that, in the present setting, the dominant correlation across the cut is captured consistently by both quantities.

However, we find that this numerical agreement in the SSH chain is not a generic equivalence. 
Note that MI fundamentally measures the \textit{total} correlations, inherently containing both quantum and classical contributions in a mixed state~\cite{Henderson2001}. As we demonstrate analytically below, using a solvable toy model, the EPE and mutual information satisfy the asymptotic relation $\mathcal{E}_B(A) \simeq I \times s(\lambda)$. Consequently, in regimes where the local entropy per channel is small [$s(\lambda) \to 0$], the EPE becomes significantly smaller than MI. 

\subsection{Toy model for the separation between EPE and mutual information}
\label{app:toy_MI_PE}
Consider $N$ independent identical two-mode Gaussian blocks, each containing one mode in $A$ and one mode in $B$, with covariance matrix
\begin{equation}
    C_{\mathrm{pair}}=
    \begin{pmatrix}
        \lambda & c\\
        c & \lambda
    \end{pmatrix},
\label{eq:toy_pair_cov}
\end{equation}
where $0<1-\lambda\ll 1$ and $|c|\ll 1$. The physicality of this Gaussian covariance matrix requires
\begin{equation}
    |\lambda\pm c|\le 1.
\end{equation}
For definiteness, we take $c\ge 0$, so that
\begin{equation}
    c\le 1-\lambda.
\end{equation}
This model is not intended as a microscopic description of a specific lattice Hamiltonian; rather, it provides a minimal analytically tractable example of weakly supported $A$-$B$ cross correlations in a mixed Gaussian setting.
For a single pair, the reduced covariance matrix on either subsystem is simply $C_A=C_B=\lambda$, and hence
\begin{equation}
    \mathcal S(A)=\mathcal S(B)=s(\lambda).
\end{equation}
The covariance matrix on the union $A\oplus B$ has eigenvalues $\lambda\pm c$, so its entropy is
\begin{equation}
    \mathcal S(A\oplus B)=s(\lambda+c)+s(\lambda-c).
\end{equation}
Expanding for small $c$ gives
\begin{equation}
    s(\lambda\pm c)
    \simeq
    s(\lambda)\pm c\,s'(\lambda)+\frac{c^2}{2}s''(\lambda),
\end{equation}
and using
\begin{equation}
    s''(x)=-\frac{1}{1-x^2},
\end{equation}
we obtain, to leading nontrivial order,
\begin{equation}
    \mathcal S(A\oplus B)
    \simeq
    2s(\lambda)-\frac{c^2}{1-\lambda^2}.
\end{equation}
Therefore the mutual information carried by a single pair is
\begin{equation}
    I_{\mathrm{pair}}
    =
    2s(\lambda)-\mathcal S(A\oplus B)
    \simeq
    \frac{c^2}{1-\lambda^2}.
\label{eq:toy_I_pair}
\end{equation}
We next evaluate the EPE for the same pair. Since $C_A=\lambda$ and $C_{AB}=c$, the directional weight defined in the main text reduces to
\begin{equation}
    w=\frac{C_{AB}C_{BA}}{1-\lambda^2}
    =
    \frac{c^2}{1-\lambda^2}.
\end{equation}
The EPE of a single pair is therefore
\begin{equation}
    \mathcal{E}_{\mathrm{pair}}
    =
    w\, s(\lambda)
    =
    \frac{c^2}{1-\lambda^2}s(\lambda).
\label{eq:toy_E_pair}
\end{equation}
Combining Eqs.~\eqref{eq:toy_I_pair} and \eqref{eq:toy_E_pair}, we find
\begin{equation}
    \mathcal{E}_{\mathrm{pair}}
    \simeq
    I_{\mathrm{pair}}\, s(\lambda).
\label{eq:toy_relation}
\end{equation}
Equation~\eqref{eq:toy_relation} shows that, relative to the mutual information, the EPE carries an additional suppression factor $s(\lambda)$. In the low-entropy regime $\lambda\to 1$, one has $s(\lambda)\to 0$, and thus $\mathcal{E}_{\mathrm{pair}}$ becomes parametrically smaller than $I_{\mathrm{pair}}$ even when both remain nonzero. In this sense, the EPE is more selective than the mutual information toward weakly supported cross correlations.
For $N$ independent copies, both quantities are additive:
\begin{equation}
    I=N\, I_{\mathrm{pair}},
    \qquad
    \mathcal{E}_B(A)=N\, \mathcal{E}_{\mathrm{pair}}.
\end{equation}
Hence
\begin{equation}
    \mathcal{E}_B(A)
    \simeq
    I\, s(\lambda).
\label{eq:toy_total_relation}
\end{equation}

\section{Finite-temperature EPE for free fermion chain by conformal mapping}

In this appendix, we provide a careful derivation of the EPE for a 1D free fermion at inverse temperature $\beta$, with subsystem $A=(0,\ell)$ and complement $B=(-\infty,0)\oplus(\ell,\infty)$. 

\subsection{Finite-temperature kernel}

For a chiral free fermion on the infinite line, the zero-temperature equal-time covariance kernel is
\begin{equation}
    C^{(0)}(y_1,y_2)=\frac{1}{2\pi i}\frac{1}{y_1-y_2},
\qquad y_1\neq y_2,
\end{equation}
with an implicit UV cutoff $a$. 
At finite temperature $\beta^{-1}$, the equal-time kernel can be obtained from the standard cylinder-to-plane conformal transformation. We use
\begin{equation}
    y(x)=\frac{v_F\beta}{2\pi}\left(1-e^{-2\pi x/v_F\beta}\right),
\qquad
    y'(x)=e^{-2\pi x/v_F\beta}.
\label{eq:map_appendix_rigorous}
\end{equation}
In the following, we use $v_F=1$ for simplicity.
This map is smooth and strictly monotone, hence it defines a bijection
\begin{equation}
    x\in\mathbb{R}
    \quad\longleftrightarrow\quad
    y\in\Omega_y:=(-\infty,\beta/2\pi).
\end{equation}
Importantly, $\Omega_y$ is not the full real line. 

By conformal covariance of a chiral fermion of scaling dimension $1/2$, the finite-temperature kernel is
\begin{equation}
    C^\beta(x_1,x_2)
    =
    \sqrt{y'(x_1)y'(x_2)}\,
    C^{(0)}\!\bigl(y(x_1),y(x_2)\bigr).
\label{eq:Cbeta_cov_rigorous}
\end{equation}

The subsystem $A=(0,\ell)$ is mapped to
\begin{equation}
    A_{\rm eff}\equiv (y(0),y(\ell))
    =
    \left(0,L_{\rm eff}\right),
\end{equation}
where as defined in the main text, $L_{\rm eff}=\frac{\beta}{2\pi}(1-e^{-2\pi\ell/\beta}).$
The complement $B=\mathbb{R}\setminus A$ is mapped to
\begin{equation}
    B_{\rm eff}:=\Omega_y\setminus A_{\rm eff}
    =
    (-\infty,0)\oplus\left(y(\ell),\beta/2\pi\right).
\end{equation}

We define the map from $\Omega_y$ to $\mathbb{R}$
\begin{equation}
    (U\psi)(x)=\sqrt{y'(x)}\,\psi(y(x)).
\label{eq:U_def_rigorous}
\end{equation}
Since $dy=y'(x)\,dx$, one has
\begin{equation}
    \int_{\mathbb{R}}dx\, |(U\psi)(x)|^2
    =
    \int_{\Omega_y}dy\, |\psi(y)|^2,
\end{equation}
so $U$ is isometry.

Restricting Eq.~\eqref{eq:Cbeta_cov_rigorous} to $A$, the covariance operator satisfies
\begin{equation}
    C_A^\beta
    =
    U_A\, C^{(0)}_{A_{\rm eff}}\, U_A^{-1},
\label{eq:C_similarity_rigorous}
\end{equation}
where $U_A$ denotes the restriction of $U$ between $L^2(A_{\rm eff})$ and $L^2(A)$.

Because Eq.~\eqref{eq:C_similarity_rigorous} is a genuine similarity transformation, any analytic functional calculus is transported accordingly. For any analytic function $f$,
\begin{equation}
    f(C_A^\beta)=U_A\, f(C^{(0)}_{A_{\rm eff}})\, U_A^{-1}.
\label{eq:functional_calc_rigorous}
\end{equation}
In particular,
\begin{equation}
    \frac{s(C_A^\beta)}{I_A-(C_A^\beta)^2}
    =
    U_A
    \frac{s(C^{(0)}_{A_{\rm eff}})}{I_{A_{\rm eff}}-(C^{(0)}_{A_{\rm eff}})^2}
    U_A^{-1}.
\label{eq:entropy_weight_rigorous}
\end{equation}

The projected kernel entering the EPE is
\begin{equation}
    K_{AB}^\beta(x_1,x_2)
    :=
    \int_B dz\, C^\beta(x_1,z)C^\beta(z,x_2).
\label{eq:K_def_rigorous}
\end{equation}
Substituting Eq.~\eqref{eq:Cbeta_cov_rigorous} to the above formula gives
\begin{align}
    K_{AB}^\beta(x_1,x_2)
    &=
    \sqrt{y'(x_1)y'(x_2)}
    \int_B dz\, y'(z)\,
    C^{(0)}(y(x_1),y(z))\nonumber
        \\ & \qquad\quad \times 
    C^{(0)}(y(z),y(x_2)).
\end{align}
Changing variables from $z$ to $\tilde y=y(z)$ and using $B\leftrightarrow B_{\rm eff}$, one finds
\begin{equation}
    K_{AB}^\beta(x_1,x_2)
    =
    \sqrt{y'(x_1)y'(x_2)}
    \int_{B_{\rm eff}} d\tilde y\,
    C^{(0)}(y(x_1),\tilde y)\,
    C^{(0)}(\tilde y,y(x_2)).
\end{equation}
Hence
\begin{equation}
    K_{AB}^\beta
    =
    U_A\, K^{(0)}_{A_{\rm eff}B_{\rm eff}}\, U_A^{-1},
\label{eq:K_similarity_rigorous}
\end{equation}
where $K^{(0)}_{A_{\rm eff}B_{\rm eff}}\equiv{}C^{(0)}_{A_{\rm eff}B_{\rm eff}}C^{(0)}_{B_{\rm eff}A_{\rm eff}}$ is defined with respect to the image-domain decomposition $\Omega_y=A_{\rm eff}\oplus B_{\rm eff}$.

\subsection{Image-domain representation and the role of thermal purification}

Using Eqs.~\eqref{eq:entropy_weight_rigorous} and
\eqref{eq:K_similarity_rigorous}, the finite-temperature entanglement projected
entropy becomes
\begin{align}
    \mathcal{E}_B(\ell,\beta)
    &=
    \Tr_A\!\left[
        K_{AB}^\beta
        \frac{s(C_A^\beta)}{I_A-(C_A^\beta)^2}
    \right]
    \nonumber\\
    &=
    \Tr_{A_{\rm eff}}\!\left[
        K^{(0)}_{A_{\rm eff}B_{\rm eff}}
        \frac{s(C^{(0)}_{A_{\rm eff}})}
        {I_{A_{\rm eff}}-(C^{(0)}_{A_{\rm eff}})^2}
    \right].
\label{eq:projected_entropy_image_domain}
\end{align}
This establishes an exact equivalence between the finite-temperature projected
entropy on the physical line and an image-domain vacuum problem.

There is, however, an important subtlety. The conformal map sends the physical line not to the full vacuum line but to
\begin{equation}
    \Omega_y=(-\infty,\beta/2\pi),
\end{equation}
It indicates that remaining part of the full vacuum line,
\begin{equation}
    E_{\rm eff}=(\beta/2\pi,\infty),
\end{equation}
is not part of the physical complement. It represents the purification sector
associated with the thermal mixedness.

Therefore, although the full vacuum covariance matrix satisfies the pure-state
identity on the complete line,
\begin{equation}
    I_{A_{\rm eff}}-(C^{(0)}_{A_{\rm eff}})^2
    =
    K^{(0)}_{A_{\rm eff}B_{\rm eff}}
    +
    K^{(0)}_{A_{\rm eff}E_{\rm eff}},
\label{eq:image_domain_purity_identity}
\end{equation}
the physical projected kernel
\(K^{(0)}_{A_{\rm eff}B_{\rm eff}}\) alone does not cancel the denominator in
Eq.~\eqref{eq:projected_entropy_image_domain}.

The projected entropy thus computes the contribution to the entropy of
\(A_{\rm eff}\) carried by the physical image complement \(B_{\rm eff}\), while
discarding the contribution carried by \(E_{\rm eff}\).

\subsection{Universal scaling}
We now evaluate the universal part of
Eq.~\eqref{eq:projected_entropy_image_domain}. The previous discussion shows
that the finite-temperature projected entropy is the image-domain contribution
carried only by the physical image complement \(B_{\rm eff}\). Therefore,
using the complement-side contour density, one should write
\begin{equation}
    \mathcal E_B(\ell,\beta)
    =
    \mathcal E_{B_{\rm eff}}(A_{\rm eff})
    =
    \int_{B_{\rm eff}} dy\, s_A(y).
\label{eq:EB_contour_general}
\end{equation}
For the interval
\begin{equation}
    A_{\rm eff}=(0,L_{\rm eff}),
\end{equation}
the full vacuum complement is
\begin{equation}
    B_{\rm full}
    =
    (-\infty,0)\oplus(L_{\rm eff},\infty).
\end{equation}
However, the physical complement of the original finite-temperature problem
maps only to
\begin{equation}
    B_{\rm eff}
    =
    (-\infty,0)
    \oplus
    \left(
        L_{\rm eff},\frac{\beta}{2\pi}
    \right).
\end{equation}
The remaining part
\begin{equation}
    E_{\rm eff}
    =
    \left(
        \frac{\beta}{2\pi},\infty
    \right)
\end{equation}
belongs to the purification sector and is not included in the physical
projected entropy. Thus
\begin{equation}
    B_{\rm full}
    =
    B_{\rm eff}\oplus E_{\rm eff},
\end{equation}
and correspondingly
\begin{equation}
    \mathcal E_{B_{\rm full}}(A_{\rm eff})
    =
    \mathcal E_{B_{\rm eff}}(A_{\rm eff})
    +
    \mathcal E_{E_{\rm eff}}(A_{\rm eff}).
\end{equation}
The quantity of interest is only
\begin{equation}
    \mathcal E_B(\ell,\beta)
    =
    \mathcal E_{B_{\rm eff}}(A_{\rm eff}),
\end{equation}
namely
\begin{equation}
    \mathcal E_B(\ell,\beta)
    =
    \int_{-\infty}^{0}dy\,s_A(y)
    +
    \int_{L_{\rm eff}}^{\beta/(2\pi)}dy\,s_A(y),
\label{eq:EB_left_right_contour}
\end{equation}
with appropriate UV cutoffs near the two endpoints of \(A_{\rm eff}\).
The discarded purification contribution is
\begin{equation}
    \mathcal E_{E_{\rm eff}}(A_{\rm eff})
    =
    \int_{\beta/(2\pi)}^\infty dy\,s_A(y).
\label{eq:discarded_E_contour}
\end{equation}

For a \((1+1)\)D CFT vacuum interval
\(A_{\rm eff}=(0,L_{\rm eff})\), the universal complement-side contour
density is
\begin{equation}
    s_A(y)
    =
    \frac{c}{6}
    \frac{L_{\rm eff}}{(-y)(L_{\rm eff}-y)},
    \qquad y<0,
\label{eq:left_contour_Leff}
\end{equation}
on the left complement, and
\begin{equation}
    s_A(y)
    =
    \frac{c}{6}
    \frac{L_{\rm eff}}{y(y-L_{\rm eff})},
    \qquad y>L_{\rm eff},
\label{eq:right_contour_Leff}
\end{equation}
on the right complement.

The UV cutoff is uniform in the original physical coordinate \(x\), with value
\(a\). Under the conformal map, it becomes position dependent in the image
coordinate:
\begin{equation}
    \delta_y(x)=y'(x)a.
\end{equation}
For the map $y(x)
    =
    \frac{\beta}{2\pi}
    \left(
        1-e^{-2\pi x/\beta}
    \right), $
one has $y'(x)=e^{-2\pi x/\beta}.$
Therefore the image cutoffs near the two endpoints of \(A_{\rm eff}\) are
\begin{equation}
    \delta_0=a,
    \qquad
    \delta_{L_{\rm eff}}
    =
    e^{-2\pi\ell/\beta}a.
\label{eq:image_cutoffs_Leff}
\end{equation}

The left contribution comes from the interval \((-\infty,0)\). With the UV
cutoff at \(y=0\), it is
\begin{align}
    \mathcal E_L
    &=
    \frac{c}{6}
    \int_{-\infty}^{-\delta_0}
    dy\,
    \frac{L_{\rm eff}}{(-y)(L_{\rm eff}-y)}
    \nonumber   =
    \frac{c}{6}
    \ln\frac{L_{\rm eff}}{a}
    +\mathrm{const}.
\label{eq:left_contribution_Leff}
\end{align}
The right physical-complement contribution comes only from $
\left(
    L_{\rm eff},\frac{\beta}{2\pi}
\right)$. Thus
\begin{align}
    \mathcal E_R^{\rm phys}
    &=
    \frac{c}{6}
    \int_{L_{\rm eff}+\delta_{L_{\rm eff}}}^{\beta/(2\pi)}
    dy\,
    \frac{L_{\rm eff}}{y(y-L_{\rm eff})}
    \nonumber\\
    &=
    \frac{c}{6}
    \ln\left[
        \frac{
        \left(\frac{\beta}{2\pi}-L_{\rm eff}\right)
        \left(L_{\rm eff}+\delta_{L_{\rm eff}}\right)
        }
        {
        \frac{\beta}{2\pi}\,
        \delta_{L_{\rm eff}}
        }
    \right]
    +\mathrm{const}\\
    &=
    \frac{c}{6}
    \ln\left[
        \frac{
        L_{\rm eff}
        \left(\frac{\beta}{2\pi}-L_{\rm eff}\right)
        }
        {
        \frac{\beta}{2\pi}\,
        \delta_{L_{\rm eff}}
        }
    \right]
    +\mathrm{const}.
\end{align}
Using $\frac{\beta}{2\pi}-L_{\rm eff}
    =
    \frac{\beta}{2\pi}
    e^{-2\pi\ell/\beta},$
the exponential factors cancel:
\begin{equation}
    \frac{
        L_{\rm eff}
        \left(\frac{\beta}{2\pi}-L_{\rm eff}\right)
    }
    {
        \frac{\beta}{2\pi}\,
        \delta_{L_{\rm eff}}
    }
    =
    \frac{L_{\rm eff}}{a}.
\end{equation}
Therefore
\begin{equation}
    \mathcal E_R^{\rm phys}
    =
    \frac{c}{6}
    \ln\frac{L_{\rm eff}}{a}
    +\mathrm{const}.
\end{equation}

Combining the left and the right physical-complement contributions, we finally obtain
\begin{equation}
    \mathcal{E}_B(\ell,\beta)
    \simeq
    \frac{c}{3}
    \ln\!\left[
        \frac{\beta}{2\pi a}
        \left(
            1-e^{-2\pi \ell/\beta}
        \right)
    \right]
    +\mathrm{const}.
\end{equation}

\end{document}